# On Magnetized Rossby Wave Dynamics in the Earth's Ionosphere

Kadri Kurt[1], Tamaz Kaladze[2,3,a], Melik Buğra Yeşil[4]

[1]*Batman University, Besiri Vocational School, 72060, Batman, Turkey*

[2]*I.Vekua Institute of Applied Mathematics, I. Javakhishvili Tbilisi State University, 0186, Georgia*

[3]*E. Andronikashvili Institute of Physics, I. Javakhishvili Tbilisi State University, 0186, Georgia*

[4]*Osmaniye Korkut Ata University, Duzici Vocational School, 80000, Osmaniye, Turkey*

**Abstract.** In the frame of shallow water model the influence of free surface action on nonlinear propagation of planetary ULF magnetized Rossby waves in the weakly ionized ionospheric D-, E-, and F-layers is revealed. Relevant nonlinear dynamic equations satisfying several conservation laws are obtained and investigated. The role of Hall and Pedersen conductivities is investigated explicitly. It is shown that while potential vorticity is conserved in the ionospheric D-, and E-layer it is broken by Pedersen conductivity in the ionospheric F-layer. Similar to KdV nonlinearity two new scalar nonlinearities due to Pedersen conductivity are revealed in the F-layer. Obtained results extend and complement known theoretical investigations and are especially relevant for nonlinear vortical propagation of magnetized Rossby waves in the weakly ionized ionospheric plasma.

**I. Introduction**

According to the experimental observations ULF (ultra-low frequency) planetary-scale perturbations (of the order of $10^3$ km) permanently exist in the Earth's ionospheric E- and F-regions (see e.g. Ref. [1-3]). Here, the distinguished role of Rossby type perturbations in the Earth's global atmospheric circulation processes and energy transfer should be emphasized [4]. Propagating at fixed latitude along the Earth's parallels Rossby waves easily form different solitary structures (e.g. vortical motions, zonal flows, etc.) at the nonlinear stage [5]. Valuable investigation of the dynamics of large-scale Rossby waves in the Earth's ionosphere needs accounting of the existence of charged particles and inhomogeneity of the turbulent state of the ionospheric gas.

Generalization of the ordinary tropospheric large-scale ULF Rossby waves and corresponding nonlinear vortical structures to the weakly ionized ionospheric D-, E-, and F-layers was performed in [6-11]. It was shown that along with the spatial inhomogeneity (along the Earth's meridians) of the Coriolis parameter, the same inhomogeneity of the geomagnetic field $\mathbf{B}_o$ should be taken into account. As a result, specific type of Rossby waves with the altered propagation phase velocity appear. Such waves are stimulated by the dynamo electric field $\mathbf{E}_d = \mathbf{v} \times \mathbf{B}_0$, and do not perturb the geomagnetic field. Some premises of the possibility of existence of planetary Rossby waves in the dynamo area of weakly ionized ionosphere was discussed by Forbes [12]. First complete theory of such waves was presented by Kaladze [8], and named magnetized Rossby waves. Further, this theory was developed in [10, 11]. Investigations [6-11] showed that interaction of induced electric current density $\mathbf{j}$ with the inhomogeneous along the meridians geomagnetic field $\mathbf{B}_o$ through the magnetic force $\mathbf{j} \times \mathbf{B}_o$ significantly modifies the dynamics of magnetized Rossby waves. The frequency of magnetized Rossby waves changes in the interval

a) Corresponding author: tamaz_kaladze@yahoo.com

$10^{-6} – 10^{-3}$ s$^{-1}$ and its wavelength is of the order of $10^3$ km and more. Phase velocity is compared to the velocity of the local winds, i.e. (1-100) m/s.

The goal of our investigation is to reveal free surface action in the shallow water model on nonlinear equations governing the dynamics of magnetized Rossby waves in the Earth's ionospheric layers. As the ionosphere is conductive medium, we involve the contribution of Hall and Pedersen conductivities in such dynamics. Note that analogous investigation for the fluid but with rigid surface under the action of shear flow was performed in [13]. As it is shown below, consideration of free surface action brings to the general nonlinear equations containing both scalar and vector nonlinearities. Inclusion of scalar nonlinearity gives the possibility to consider monopolar vortical structures while in Ref. [13], there is no such possibility. In addition inclusion of scalar nonlinearity reveals the appearance of two new scalar nonlinearities in the ionospheric F-layer which were unknown still.

The given presentation is organized as follows: in the section 2, based on the shallow water model equations modeling the nonlinear propagation of the magnetized Rossby waves in the ionosphere are given. Necessary for development small parameters (see below Eqs. (2.9) - (2.11)) are discussed. In the section 3, we discuss potential vorticity conservation problem in different layers of the ionosphere. In the section 4, we obtain nonlinear vorticity equations describing the dynamics of magnetized Rossby waves in the Earth's ionosphere. It is shown that such equations satisfy necessary conservation laws. Note that the obtained equations are general and are valid for the wide heights (50 – 500 km) of the ionosphere. Discussion and conclusion are given in the section 5.

**II. Initial Equations**

We examine propagation dynamics of large-scale planetary ULF magnetized Rossby waves in the Earth's weakly ionized ionospheric D-, E-, and F-layers formed by electrons, ions and bulk of neutral particles, so the inequality n/N<<1 (n and N are the concentrations of the charged and neutral particles, respectively) is fulfilled. It follows that the dynamics of such ionospheric medium is stimulated by the neutral component. Charged particles are involved in the dynamics of such weakly ionized gas due to the strong collision coupling between the ionized components with the neutral component. Experimental observations [3, 14-16] indicate that the large-scale motions in the ionosphere are almost horizontal due to the smallness of vertical velocity compared to horizontal one.

Let us introduce the local Cartesian system of coordinates $(x,y,z)$, where $x$ and $y$ are directed eastward (along the parallel) and northward (along the meridian), respectively, and vertical $z-$axis is directed upward. In addition, we consider 2D magnetized Rossby waves on the $(x,y)$ plane with the horizontal velocity $\mathbf{v}(u,v,0)$ and ignore the derivative $\partial/\partial z = 0$ for perturbed quantities. For the set problem we consider a shallow fluid model with the assumption that a perturbation of the free surface occurs (i.e. the two-dimensional divergence of the fluid velocity is not equal to zero) [15, 16]. In addition, we assume that mass density of the fluid $\rho = const$. Thus, our initial equations consist of the following neutral gas momentum equation and the corresponding incompressibility condition (cf. [13])

$$\frac{\partial \mathbf{v}}{\partial t} + (\mathbf{v}\cdot \nabla)\mathbf{v} = -g\nabla H + 2\mathbf{v}\times \mathbf{\Omega} + \frac{1}{\rho}\mathbf{j}\times \mathbf{B_0} \, , \tag{2.1}$$

$$\frac{\partial u}{\partial x} + \frac{\partial v}{\partial y} = -\frac{d}{dt}(\ln H) \quad , \tag{2.2}$$

where **v** is the velocity of the incompressible neutral gas, $\mathbf{\Omega}$ is the angular velocity of the Earth's rotation, $\rho = NM$ is the density of the medium, $g$ is the gravitational acceleration, the total depth of the fluid $H = H_0 + \delta H(t, x, y) \equiv H_0(1+h)$, $H_0$ is the uniform thickness of fluid in its stationary state, $\delta H$ is the deviation of free surface from $H_0$ and $h = \delta H / H_0 < 1$. Acting forces in Eq. (2.1) are presented by the action of free surface, Coriolis, and magnetic forces. In the chosen coordinate system the geomagnetic induction $\mathbf{B}_0$ and the Earth's angular velocity $\mathbf{\Omega}$ have the components $\mathbf{B}_0(0, B_{0y}, B_{0z})$, $\mathbf{\Omega}(0, \Omega_y, \Omega_z)$ and both do not depend on $x$ – coordinate but on latitudinal $y$ - coordinate.

We will use the generalized Ohm's law to describe electric conductivity of the weakly ionized ionospheric layers [2]

$$\mathbf{j} = \sigma_{\parallel}\mathbf{E}_{d\parallel} + \sigma_{\perp}\mathbf{E}_{d\perp} + \sigma_H \frac{\mathbf{B}_0 \times \mathbf{E}_d}{B_0}, \tag{2.3}$$

where $\mathbf{E}_d = \mathbf{v} \times \mathbf{B}_0$ is the dynamo electric field. $\parallel$, and $\perp$ indicate parallel and perpendicular components, respectively. We use induction-free approximation, when the magnetic field generated by the electric current density $\mathbf{j}$ can be ignored, so the total induction $\mathbf{B}$ will be equal to the external geomagnetic field $\mathbf{B}_0$.

In Eq. (2.3) $\sigma_{\parallel}$ is the parallel conductivity (in the direction of the geomagnetic field), $\sigma_p = \sigma_{\perp}$ is the Pedersen (transverse to the geomagnetic field) conductivity, and $\sigma_H$ is the Hall conductivity, which are defined as follows [17]

$$\sigma_{\parallel} = \frac{ne^2}{m_e \nu_e}, \tag{2.4}$$

$$\sigma_p = \sigma_{\perp} = \frac{ne^2 \nu_{in}(\nu_e \nu_{in} + \omega_{ce}\omega_{ci})}{m_e(\omega_{ce}^2 \nu_{in}^2 + \nu_e^2 \nu_{in}^2 + \omega_{ce}^2 \omega_{ci}^2)}, \tag{2.5}$$

$$\sigma_H = \frac{ne^2 \nu_{in}^2 \omega_{ce}}{m_e(\omega_{ce}^2 \nu_{in}^2 + \nu_e^2 \nu_{in}^2 + \omega_{ce}^2 \omega_{ci}^2)}. \tag{2.6}$$

Here $\nu_{en}$ and $\nu_{in}$ are the effective collisional frequencies of the neutral particles with the electrons and ions, $e$ is the electron charge, $\nu_e = \nu_{ei} + \nu_{en}$, $\nu_{ei}$ is the effective collisional frequency of electrons with ions, $\omega_{ce} = eB_0/m_e$ and $\omega_{ci} = ZeB_0/m_i$ are the cyclotron frequencies of electrons and ions, $m_e$ and $m_i$ are the electron and ion masses, and $Z$ is the ion charge state.

Under the described assumptions, we represent Eq. (2.1) as follows (see more details in [13])

$$\frac{\partial \mathbf{v}}{\partial t} + (\mathbf{v} \cdot \nabla)\mathbf{v} = -g\nabla H + 2\mathbf{v} \times \mathbf{\Omega} - \frac{1}{\rho}\sigma_{\perp}B_0^2 \mathbf{v} + \frac{1}{\rho}\sigma_{\perp}\mathrm{v}_y B_{0y}\mathbf{B}_0 + \frac{1}{\rho}\sigma_H B_0 \mathbf{v} \times \mathbf{B}_0. \tag{2.7}$$

Further, we will use the $\beta$-plane approximation and represent

$$f(y) = 2\Omega_z(y) = f_0 + \beta y, \quad \gamma(y) = \sigma_H B_0 B_{0z}/\rho = \gamma_0 + \alpha y, \tag{2.8}$$

where $\beta = \partial f \partial / \partial y$, $\alpha = \partial \gamma / \partial y$, so $f_0$ and $\gamma_0$ are the relevant values at $y = 0$.

Introducing the typical length and velocity scales $L$, $U$ in what follows we will use the following scaling parameters

1) $\varepsilon_\omega = \dfrac{U}{fL} \sim \dfrac{\omega}{f} \sim \dfrac{\zeta_z}{f} \leq 1$, which is the Rossby number estimation (2.9)

2) $\varepsilon_\beta = \dfrac{L\beta}{f} = \dfrac{L\partial f/\partial y}{f} \sim \dfrac{L}{R} \ll 1$, (2.10)

3) $\varepsilon_h = h \sim \dfrac{UL}{fr_R^2} \ll 1$ . (2.11)

Here $\omega$ is the typical frequency of Rossby waves, $f = 2\Omega_z$ is the Coriolis parameter, $\zeta_z$ is the z-component of the vorticity $\zeta_z = \hat{\mathbf{z}}\cdot\nabla\times\mathbf{v}$, $\beta = \partial f/\partial y$ is the Rossby parameter, $R$ is the Earth's radius, $r_R = \sqrt{gH_0}/f$ is the Rossby radius. Equation (2.10) represents the principle parameter restriction for the validity of the $\beta$ – plane approximation.

### III. Potential vorticity

From Eq. (2.7), we find the following relation for the z-component of vorticity

$$\dfrac{\partial \zeta_z}{\partial t} + u\dfrac{\partial \zeta_z}{\partial x} + v\dfrac{\partial \zeta_z}{\partial y} + \zeta_z(\dfrac{\partial u}{\partial x} + \dfrac{\partial v}{\partial y}) =$$
$$-f(\dfrac{\partial u}{\partial x} + \dfrac{\partial v}{\partial y}) - \dfrac{\sigma_\perp B_0^2}{\rho}\zeta_z - \beta v - \dfrac{\sigma_H B_0 B_{0z}}{\rho}(\dfrac{\partial u}{\partial x} + \dfrac{\partial v}{\partial y}) + \dfrac{\sigma_\perp B_{0y}^2}{\rho}\dfrac{\partial v}{\partial x} +$$
$$u\dfrac{\partial}{\partial y}(\dfrac{\sigma_\perp B_0^2}{\rho}) - v\dfrac{\partial}{\partial y}(\dfrac{\sigma_H B_0 B_{0z}}{\rho})$$
(3.1)

Now we use Eq. (2.2), then Eq. (3.1) becomes

$$\dfrac{d}{dt}\zeta_z - \zeta_z\dfrac{1}{H}\dfrac{dH}{dt} = f\dfrac{1}{H}\dfrac{dH}{dt} - \dfrac{1}{\rho}\sigma_\perp B_0^2 \zeta_z - \dfrac{df}{dt} + \dfrac{1}{\rho}\sigma_H B_0 B_{0z}\dfrac{1}{H}\dfrac{dH}{dt}$$
$$+ \dfrac{1}{\rho}\sigma_\perp B_{0y}^2\dfrac{\partial v}{\partial x} + u\dfrac{\partial}{\partial y}(\dfrac{\sigma_\perp B_0^2}{\rho}) - \dfrac{d}{dt}(\dfrac{1}{\rho}\sigma_H B_0 B_{0z})$$
(3.2)

Here the operator $\dfrac{d}{dt} = \dfrac{\partial}{\partial t} + u\dfrac{\partial}{\partial x} + v\dfrac{\partial}{\partial y}$ is the total derivative along the particle's trajectory. Eq. (3.2) can be rewritten as

$$\dfrac{d}{dt}(\dfrac{\zeta_z + f + \gamma}{H}) = \dfrac{1}{\rho H}\dfrac{\partial}{\partial y}(u\sigma_\perp B_0^2) - \dfrac{1}{\rho H}\sigma_\perp B_{0z}^2\dfrac{\partial v}{\partial x} \quad .$$
(3.3)

a) Let us consider the ionospheric D-layer ($50-90\, km$ heights from the Earth's surface). As it was shown in [13], ionospheric D-layer might be considered as un-magnetized, i.e. conductivity contributions could be ignored. Thus, we get

$$\frac{d}{dt}\Pi = 0,\tag{3.4}$$

where the potential vorticity

$$\Pi = \frac{\zeta_z + f}{H}\tag{3.5}$$

Is conserved.

b) In the Ionospheric E-layer ($90-160\,km$ heights from the Earth's surface), Hall conductivity prevails Pedersen conductivity contribution [13], i.e. Eq. (3.3) modifies as follows

$$\frac{d}{dt}\Pi = 0,\tag{3.6}$$

where the potential vorticity in the geomagnetic field

$$\Pi = \frac{\zeta_z + f + \dfrac{\sigma_H B_0 B_{0z}}{\rho}}{H}\tag{3.7}$$

Is conserved.

c) In the ionospheric F-layer ($160-500\,km$ heights from the Earth's surface) Eq. (3.3) yields

$$\frac{d}{dt}\Pi = \frac{1}{H}\frac{\partial}{\partial y}(u\frac{\sigma_\perp B_0^2}{\rho}) - \frac{1}{H}\frac{\sigma_\perp B_{0z}^2}{\rho}\frac{\partial v}{\partial x},\tag{3.8}$$

where $\Pi$ is expressed by Eq. (3.7).

We see that in the ionospheric E-layer with $\sigma_H \gg \sigma_\perp$, potential vorticity is conserved, while in the ionospheric F-layer with $\sigma_\perp \gg \sigma_H$ potential vorticity is not conserved.

**IV. Nonlinear propagation dynamics**

Now we consider low-frequency branch of Rossby type waves, which is characterized by a small Rossby number (2.9) and in the geostrophic equilibrium we obtain from Eq. (2.7) the following system

$$\begin{cases} -g\dfrac{\partial(\delta H)}{\partial x} + v(f + \dfrac{\sigma_H B_0 B_{0z}}{\rho}) - \dfrac{\sigma_\perp B_0^2}{\rho}u = 0, \\ -g\dfrac{\partial(\delta H)}{\partial y} - u(f + \dfrac{\sigma_H B_0 B_{0z}}{\rho}) - v\dfrac{\sigma_\perp B_{0z}^2}{\rho} = 0 \end{cases}\tag{4.1}$$

Solutions are

$$\begin{cases} u = -\dfrac{Ag\dfrac{\partial(\delta H)}{\partial y} + g\dfrac{\partial(\delta H)}{\partial x}\dfrac{\sigma_\perp B_{0z}^2}{\rho}}{A^2 + (\dfrac{\sigma_\perp B_{0z} B_0}{\rho})^2}, \\ \\ v = \dfrac{Ag\dfrac{\partial(\delta H)}{\partial x} - g\dfrac{\partial(\delta H)}{\partial y}\dfrac{\sigma_\perp B_0^2}{\rho}}{A^2 + (\dfrac{\sigma_\perp B_{0z} B_0}{\rho})^2}. \end{cases} \quad (4.2)$$

Here

$$A = f + \frac{1}{\rho}\sigma_H B_0 B_{0z}, \qquad B_0^2 = B_{0y}^2 + B_{0z}^2 . \quad (4.3)$$

From Eq. (3.2), we obtain the following vorticity equation valid for the ionospheric D-, E-, and F- layers

$$(1+h)\frac{d\zeta_z}{dt} - (\zeta_z + f + \gamma)\frac{dh}{dt} + (1+h)\frac{d}{dt}(f+\gamma) = \\ -\frac{\sigma_\perp B_0^2}{\rho}(1+h)\zeta_z + \frac{\sigma_\perp B_{0y}^2}{\rho}(1+h)\frac{\partial v}{\partial x} + (1+h)u\frac{\partial}{\partial y}\left(\frac{\sigma_\perp B_0^2}{\rho}\right) \quad (4.4)$$

Here $\gamma = \sigma_H B_0 B_{0z} / \rho$ and $h = \delta H / H_0$ is the normalized displacement of the free surface.

Obtained in the previous section results may be applied to the different ionospheric D-, E-, and F-layers, which are surrounded by the geomagnetic field $\mathbf{B}_0 = 0.5 \times 10^{-4} T$. Necessary numerical values for this layers are given in the table [13].

### IV.1 Ionospheric neutral D-layer

Let us start the investigation with the ionospheric D-layer, which situates at the $50-90$ km heights from the Earth's surface. Based on numerical values of parameters, it was shown in [13], that D-layer might be considered as un-magnetized, i.e. conductivity contributions could be ignored. Thus, we may replace Eq. (4.2) – (4.4) by

$$A = f, \quad u = -\frac{gH_0}{f}\frac{\partial h}{\partial y}, \quad v = \frac{gH_0}{f}\frac{\partial h}{\partial x}, \quad \zeta_z = \frac{gH_0}{f}\Delta h, \quad (4.5)$$

$$(1+h)\frac{d\zeta_z}{dt} - (\zeta_z + f)\frac{dh}{dt} + (1+h)\frac{df}{dt} = 0. \quad (4.6)$$

We can show that, $\dfrac{dh}{dt} = \dfrac{\partial h}{\partial t}$, and taking into account that $f$ does not depend on the $x-$ coordinate, finally we obtain the following vorticity equation for the neutral D-layer [18]

$$\frac{\partial}{\partial t}\Delta h - \frac{1}{r_R^2}\frac{\partial h}{\partial t} + \beta\frac{\partial h}{\partial x} + \beta h\frac{\partial h}{\partial x} + \frac{gH_0}{f}J(h,\Delta h) = 0. \quad (4.7)$$

Here the Jacobian $J(a,b) = \partial a/\partial x \, \partial b/\partial y - \partial a/\partial y \, \partial b/\partial x$ is introduced, and $r_R = \sqrt{gH_0}/f$ is the Rossby radius. Note that along with vector nonlinearity $J(h, \Delta h)$, in Eq. (4.7) scalar $h\partial h/\partial x$ nonlinearity, Korteweg-de Vries (KdV) type nonlinearity also exists.

Let us consider some conservation laws satisfied by Eq. (4.7).

a) Integration of Eq. (4.7) by parts with the zero conditions at infinity gives the following mass conservation law

$$\frac{\partial}{\partial t} \int h \, dx dy = 0. \tag{4.8}$$

b) Multiplying Eq. (4.7) by $h$ and integration by parts gives the following energy conservation law

$$\frac{\partial}{\partial t} \int \varepsilon \, dx dy = 0, \tag{4.9}$$

where

$$\varepsilon = \frac{1}{2} h^2 + \frac{1}{2} r_R^2 (\nabla h)^2 \tag{4.10}$$

Is the local energy density, and $(\nabla h)^2 = (\partial h/\partial x)^2 + (\partial h/\partial y)^2$.

c) Let us multiply Eq. (4.7) by the variable $y$; integration by parts gives the following conservation law for the $y$-coordinate for the center of mass

$$\frac{\partial}{\partial t} \int y h \, dx dy = 0. \tag{4.11}$$

d) Multiplying Eq. (4.7) by $\Delta h$ and integration by parts gives the following relation for the potential enstrophy

$$\frac{\partial}{\partial t} \int dx dy [(\nabla h)^2 + r_R^2 (\Delta h)^2] = -\beta r_R^2 \int dx dy \Delta h \frac{\partial h^2}{\partial x}, \tag{4.12}$$

where the integrand

$$q = (\nabla h)^2 + r_R^2 (\Delta h)^2 \tag{4.13}$$

is the local potential enstrophy. Thus, the total potential enstrophy $Q = \int dx dy \, q = \langle q \rangle = \langle (\nabla h)^2 + r_R^2 (\Delta h)^2 \rangle$ is not conserved due to the scalar nonlinearity. But, we should take into account that the right-hand side of Eq. (4.12) is small as cubic $O(h^3)$.

### IV.2 Ionospheric weakly ionized E-layer

Let us study the ionospheric E-layer situated at the (90-160) km heights from the Earth's surface. Suitable numerical parameters can be found in [13].

Consider relatively low heights, with $\nu_{in} \gg \omega_{ci}$, from Eqs. (2.5), (2.6), we find

$$\sigma_\perp = \frac{ne^2}{m_e} \frac{1}{\omega_{ce}} \frac{\omega_{ci}}{\nu_{in}} \ll \sigma_H = \frac{ne^2}{m_e} \frac{1}{\omega_{ce}}. \tag{4.14}$$

Taking into account the charge particles numerical interval in the E-layer, we may use the estimation $\sigma_H = 2.9 \times (10^{-6} - 10^{-4}) S/m$. Correspondingly, using $\omega_{ci}/\nu_{in} = 2 \times 10^{-2} - 10^{-1}$ we obtain $\sigma_\perp = 2.9 \times (2 \times 10^{-8} - 10^{-5}) S/m$. In addition, the mass density changes in the numerical interval $\rho = (10^{-8} - 10^{-6}) kg/m^3$ and for the Pedersen and Hall parameters we get (considering the middle latitudes $\Omega_z \neq 0$)

$$\frac{\sigma_\perp B_{0z}^2}{\rho} \sim 0.7 \times (2 \times 10^{-10} - 10^{-5}) s^{-1} \ll 2\Omega_z, \tag{4.15}$$

$$\frac{\sigma_H |B_{0z}| B_{0z}}{\rho} \sim \frac{\sigma_H B_{0z}^2}{\rho} \sim 0.7 \times (10^{-8} - 10^{-4}) s^{-1}. \tag{4.16}$$

For the minimal value $\rho = 10^{-8} kg/m^3$ and maximal value $n = 10^{11} m^{-3}$ Pedersen parameter $\sigma_\perp B_{0z}^2/\rho = 0.7 \times 10^{-5} s^{-1}$, and Hall parameter $\sigma_H |B_{0z}| B_{0z}/\rho = 0.7 \times 10^{-4} s^{-1}$. Thus, Hall parameter can be close to the Coriolis parameter $2\Omega_z = 10^{-4} s^{-1}$, while Pedersen parameter remains smaller. Nevertheless, we keep both contributions, because the Pedersen parameter is responsible for the small magnetic damping of the magnetized Rossby waves [6-9].

Thus, in Eq. (4.4) we ignore the term $\sigma_\perp B_{0y}^2/\rho$ (considering sufficiently high latitudes) and according to the estimations (2.9), (2.10) the last term also. Finally the vorticity equation for the ionospheric E-layer is

$$(1+h)\frac{d\zeta_z}{dt} - (\zeta_z + f + \gamma)\frac{dh}{dt} + (1+h)\frac{d}{dt}(f+\gamma) = -\frac{\sigma_\perp B_0^2}{\rho}(1+h)\zeta_z \tag{4.17}$$

Here $\gamma = \sigma_H B_0 B_{0z}/\rho \sim f$. Thus, instead (4.5) we get

$$A = f + \gamma, \quad u = -\frac{gH_0}{f_0 + \gamma_0}\frac{\partial h}{\partial y}, \quad v = \frac{gH_0}{f_0 + \gamma_0}\frac{\partial h}{\partial x}, \quad \zeta_z = \frac{gH_0}{f_0 + \gamma_0}\Delta h, \quad \frac{dh}{dt} = \frac{\partial h}{\partial t}. \tag{4.18}$$

Then the vorticity equation (4.17) becomes (cf. [10, 11])

$$\frac{\partial}{\partial t}(\Delta h - \frac{1}{r_R^2}h) + (\alpha+\beta)\frac{\partial h}{\partial x} + (\alpha+\beta)h\frac{\partial h}{\partial x} + \frac{gH_0}{f_0+\gamma_0}J(h,\Delta h) + \frac{\sigma_\perp B_0^2}{\rho}\Delta h = 0 \tag{4.19}$$

Here $r_R = \sqrt{gH_0}/|f_0 + \gamma_0|$ is the Rossby radius modified by the geomagnetic field. Note that $\alpha$ and $\beta$ have opposite signs [8, 10, 11]. When deriving Eq. (4.19) we used the estimation (2.10), and ignored cubic terms $h^3$, along with $h\partial\Delta h/\partial t, \Delta h\partial/\partial t$ (see [10, 13]).

Eq. (4.19) conserves mass density (see Eq. (4.8)), and the $y$-coordinate for the center of mass (see Eq. (4.11)). As to the energy, instead of Eq. (4.9) we get

$$\frac{\partial}{\partial t}\int \varepsilon dxdy = -\frac{\sigma_\perp B_0^2}{\rho}r_R^2\int dxdy(\nabla h)^2 . \qquad (4.20)$$

Thus, the total energy in the E-layer is not conserved and damped due to the Pedersen conductivity. As to the potential enstrophy instead of Eq. (4.12) we get

$$\frac{\partial}{\partial t}\int dxdy[(\nabla h)^2 + r_R^2(\Delta h)^2] = -2r_R^2\frac{\sigma_\perp B_0^2}{\rho}\int dxdy(\Delta h)^2 - (\alpha+\beta)r_R^2\int dxdy\Delta h\frac{\partial h^2}{\partial x} . \qquad (4.21)$$

Thus, the potential enstrophy $Q = \langle(\nabla h)^2 + r_R^2(\Delta h)^2\rangle$ is not conserved due to Pedersen conductivity and scalar nonlinearity. Note that the last term on the right-hand side is small as $O(h^3)$.

**IV.3 Ionospheric weakly ionized F-layer**

Ionospheric F-layer is situated at the $(160-500)km$ heights from the Earth's surface. The suitable numerical parameters can be found in [13]. Then we find from Eqs. (2.5) and (2.6)

$$\sigma_\perp = \frac{ne^2}{m_e}\frac{1}{\omega_{ce}}\frac{\nu_{in}}{\omega_{ci}} \gg \sigma_H = \frac{ne^2}{m_e}\frac{1}{\omega_{ce}}(\frac{\nu_{in}}{\omega_{ci}})^2 , \qquad (4.22)$$

where $\sigma_\perp = (5\times10^{-8} - 6\times10^{-6})S/m$, and $\sigma_H = \sigma_\perp\frac{\nu_{in}}{\omega_{ci}} = 8.5\times10^{-12} - 1.2\times10^{-8} S/m.$

Correspondingly

$$\frac{\sigma_\perp B_{0z}^2}{\rho} = (1.25\times10^{-6} - 1.5\times10^{-4})s^{-1}, \qquad (4.23)$$

$$\frac{\sigma_H|B_{0z}|B_{0z}}{\rho} = (2\times10^{-10} - 3\times10^{-7})s^{-1} \ll 2\Omega_z = 1.5\times10^{-4}s^{-1}. \qquad (4.24)$$

Then assuming that $f \sim \sigma_\perp B_0^2/\rho$, from Eqs. (4.2) and (4.3) we find

$$\begin{cases} u = -\dfrac{gH_0}{f^2 + (\dfrac{\sigma_\perp B_0 B_{0z}}{\rho})^2}(f\dfrac{\partial h}{\partial y} + \dfrac{\sigma_\perp B_{0z}^2}{\rho}\dfrac{\partial h}{\partial x}), \\ \\ v = \dfrac{gH_0}{f^2 + (\dfrac{\sigma_\perp B_0 B_{0z}}{\rho})^2}(f\dfrac{\partial h}{\partial x} - \dfrac{\sigma_\perp B_0^2}{\rho}\dfrac{\partial h}{\partial y}) \end{cases} \qquad (4.25)$$

Using the estimation (2.9), we find

$$\zeta_z = r_R^2(f\Delta h - \frac{\sigma_\perp B_{0y}^2}{\rho}\frac{\partial^2 h}{\partial x\partial y}) , \qquad (4.26)$$

$$\frac{dh}{dt} = \frac{\partial h}{\partial t} - r_R^2 [\frac{\sigma_\perp B_{0z}^2}{\rho}(\frac{\partial h}{\partial x})^2 + \frac{\sigma_\perp B_0^2}{\rho}(\frac{\partial h}{\partial y})^2] \qquad (4.27)$$

$$\frac{d}{dt}\zeta_z = r_R^2 f \frac{\partial \Delta h}{\partial t} - r_R^2 \frac{\sigma_\perp B_{0y}^2}{\rho} \frac{\partial^3 h}{\partial t \partial x \partial y} + r_R^4 f^2 J(h, \Delta h) + r_R^4 f \frac{\partial f}{\partial y}\frac{\partial h}{\partial x}\Delta h$$
$$-r_R^4 \frac{\sigma_\perp B_0^2}{\rho} \frac{\partial f}{\partial y}\frac{\partial h}{\partial y}\Delta h + r_R^4 \frac{\sigma_\perp B_{0z}^2}{\rho}\frac{\sigma_\perp B_{0y}^2}{\rho}\frac{\partial h}{\partial x}\frac{\partial^3 h}{\partial x^2 \partial y}$$
$$+r_R^4 \frac{\sigma_\perp B_0^2}{\rho}\frac{\sigma_\perp B_{0y}^2}{\rho}\frac{\partial h}{\partial y}\frac{\partial^3 h}{\partial x \partial y^2} + r_R^4 f \frac{\sigma_\perp B_{0y}^2}{\rho}\frac{\partial h}{\partial y}\frac{\partial^3 h}{\partial x^2 \partial y} \qquad (4.28)$$
$$-r_R^4 f \frac{\sigma_\perp B_{0y}^2}{\rho}\frac{\partial h}{\partial x}\frac{\partial^3 h}{\partial x \partial y^2} - r_R^4 f \frac{\sigma_\perp B_{0z}^2}{\rho}\frac{\partial h}{\partial x}\frac{\partial \Delta h}{\partial x} - r_R^4 f \frac{\sigma_\perp B_0^2}{\rho}\frac{\partial h}{\partial y}\frac{\partial \Delta h}{\partial y}$$
$$-r_R^4 f \frac{\partial h}{\partial x}\frac{\partial^2 h}{\partial x \partial y}\frac{\partial}{\partial y}(\frac{\sigma_\perp B_{0y}^2}{\rho}) + r_R^4 \frac{\sigma_\perp B_0^2}{\rho}\frac{\partial h}{\partial y}\frac{\partial^2 h}{\partial x \partial y}\frac{\partial}{\partial y}(\frac{\sigma_\perp B_{0y}^2}{\rho})$$

$$h\frac{d\zeta_z}{dt} = r_R^2 fh\frac{\partial \Delta h}{\partial t} - r_R^2 \frac{\sigma_\perp B_{0y}^2}{\rho} h\frac{\partial^3 h}{\partial t \partial x \partial y} + O(h^3) \qquad (4.29)$$

$$\zeta_z \frac{dh}{dt} = r_R^2 f \Delta h \frac{\partial h}{\partial t} - r_R^2 \frac{\sigma_\perp B_{0y}^2}{\rho}\frac{\partial h}{\partial t}\frac{\partial^2 h}{\partial x \partial y} + O(h^3) \qquad (4.30)$$

Here, in Eqs. (4.26) - (4.30)

$$r_R^2 = \frac{gH_0}{f^2 + (\frac{\sigma_\perp B_{0z} B_0}{\rho})^2} \qquad (4.31)$$

Is modified by Pedersen conductivity Rossby radius.

Thus Eq. (4.4) for F-layer ($\gamma = \sigma_H B_0 B_{0z} / \rho = 0$) becomes

$$(1+h)\frac{d\zeta_z}{dt} - (\zeta_z + f)\frac{dh}{dt} + (1+h)\frac{df}{dt} + \frac{\sigma_\perp B_0^2}{\rho}(1+h)\zeta_z$$
$$-\frac{\sigma_\perp B_{0y}^2}{\rho}(1+h)\frac{\partial v}{\partial x} - (1+h)u\frac{\partial}{\partial y}(\frac{\sigma_\perp B_0^2}{\rho}) = 0 \qquad (4.32)$$

Now we represent remaining in Eq. (4.32) terms as follows

$$f\frac{dh}{dt} \to f_0 \frac{dh}{dt} = f_0[\frac{\partial h}{\partial t} - r_R^2 \frac{\sigma_\perp B_{0z}^2}{\rho}(\frac{\partial h}{\partial x})^2 - r_R^2 \frac{\sigma_\perp B_0^2}{\rho}(\frac{\partial h}{\partial y})^2], \qquad (4.33)$$

$$(1+h)\frac{df}{dt} = (1+h)v\frac{\partial f}{\partial y} = (1+h)\beta r_R^2 (f\frac{\partial h}{\partial x} - \frac{\sigma_\perp B_0^2}{\rho}\frac{\partial h}{\partial y}), \qquad (4.34)$$

$$\frac{1}{\rho}\sigma_\perp B_0^2(1+h)\zeta_z = \frac{1}{\rho}\sigma_\perp B_0^2 r_R^2 f \Delta h - \frac{1}{\rho}\sigma_\perp B_0^2 r_R^2 \frac{\sigma_\perp B_{0y}^2}{\rho}\frac{\partial^2 h}{\partial x \partial y}$$
$$+\frac{1}{\rho}\sigma_\perp B_0^2 r_R^2 fh\Delta h - \frac{1}{\rho}\sigma_\perp B_0^2 r_R^2 \frac{\sigma_\perp B_{0y}^2}{\rho} h\frac{\partial^2 h}{\partial x \partial y} \quad , \tag{4.35}$$

$$\frac{1}{\rho}\sigma_\perp B_{0y}^2(1+h)\frac{\partial v}{\partial x} = \frac{1}{\rho}\sigma_\perp B_{0y}^2 r_R^2 f \frac{\partial^2 h}{\partial x^2} - \frac{1}{\rho}\sigma_\perp B_{0y}^2 r_R^2 \frac{\sigma_\perp B_0^2}{\rho}\frac{\partial^2 h}{\partial x \partial y}$$
$$+\frac{1}{\rho}\sigma_\perp B_{0y}^2 r_R^2 fh\frac{\partial^2 h}{\partial x^2} - \frac{1}{\rho}\sigma_\perp B_{0y}^2 r_R^2 \frac{\sigma_\perp B_0^2}{\rho} h\frac{\partial^2 h}{\partial x \partial y} \quad , \tag{4.36}$$

$$(1+h)u\frac{\partial}{\partial y}(\frac{\sigma_\perp B_0^2}{\rho}) = -r_R^2 \frac{\partial}{\partial y}(\frac{\sigma_\perp B_0^2}{\rho})f\frac{\partial h}{\partial y} - r_R^2 \frac{\partial}{\partial y}(\frac{\sigma_\perp B_0^2}{\rho})\frac{\sigma_\perp B_{0z}^2}{\rho}\frac{\partial h}{\partial x}$$
$$-r_R^2 \frac{\partial}{\partial y}(\frac{\sigma_\perp B_0^2}{\rho})fh\frac{\partial h}{\partial y} - r_R^2 \frac{\partial}{\partial y}(\frac{\sigma_\perp B_0^2}{\rho})\frac{\sigma_\perp B_{0z}^2}{\rho} h\frac{\partial h}{\partial x} \quad . \tag{4.37}$$

Finally, vorticity Eq. (4.32) becomes

$$\frac{\partial \Delta h}{\partial t} - \frac{\sigma_\perp B_{0y}^2}{\rho f_0}\frac{\partial^3 h}{\partial t \partial x \partial y} + r_R^2 f_0 J(h,\Delta h) + r_R^2 \beta \frac{\partial h}{\partial x}\Delta h - r_R^2 \frac{\sigma_\perp B_0^2}{\rho f_0}\beta \frac{\partial h}{\partial y}\Delta h$$
$$+r_R^2 \frac{\sigma_\perp B_{0z}^2}{\rho f_0}\frac{\sigma_\perp B_{0y}^2}{\rho}\frac{\partial h}{\partial x}\frac{\partial^3 h}{\partial x^2 \partial y} + r_R^2 \frac{\sigma_\perp B_0^2}{\rho f_0}\frac{\sigma_\perp B_{0y}^2}{\rho}\frac{\partial h}{\partial y}\frac{\partial^3 h}{\partial x \partial y^2} + r_R^2 \frac{\sigma_\perp B_{0y}^2}{\rho}\frac{\partial h}{\partial y}\frac{\partial^3 h}{\partial x^2 \partial y}$$
$$-r_R^2 \frac{\sigma_\perp B_{0y}^2}{\rho}\frac{\partial h}{\partial x}\frac{\partial^3 h}{\partial x \partial y^2} - r_R^2 \frac{\sigma_\perp B_{0z}^2}{\rho}\frac{\partial h}{\partial x}\frac{\partial \Delta h}{\partial x} - r_R^2 \frac{\sigma_\perp B_0^2}{\rho}\frac{\partial h}{\partial y}\frac{\partial \Delta h}{\partial y}$$
$$-r_R^2 \frac{\partial h}{\partial x}\frac{\partial^2 h}{\partial x \partial y}\frac{\partial}{\partial y}(\frac{\sigma_\perp B_{0y}^2}{\rho}) + r_R^2 \frac{\sigma_\perp B_0^2}{\rho f_0}\frac{\partial h}{\partial y}\frac{\partial^2 h}{\partial x \partial y}\frac{\partial}{\partial y}(\frac{\sigma_\perp B_{0y}^2}{\rho})$$
$$+h\frac{\partial \Delta h}{\partial t} - \Delta h\frac{\partial h}{\partial t} - \frac{\sigma_\perp B_{0y}^2}{\rho f_0}h\frac{\partial^3 h}{\partial t \partial x \partial y} - \frac{1}{r_R^2}\frac{\partial h}{\partial t}$$
$$+\frac{\sigma_\perp B_{0y}^2}{\rho f_0}\frac{\partial h}{\partial t}\frac{\partial^2 h}{\partial x \partial y} + \frac{\sigma_\perp B_{0z}^2}{\rho}(\frac{\partial h}{\partial x})^2 + \frac{\sigma_\perp B_0^2}{\rho}(\frac{\partial h}{\partial y})^2 + \beta \frac{\partial h}{\partial x} - \beta \frac{\sigma_\perp B_0^2}{\rho f_0}\frac{\partial h}{\partial y}$$
$$+\beta h\frac{\partial h}{\partial x} - \beta \frac{\sigma_\perp B_0^2}{\rho f_0}h\frac{\partial h}{\partial y} + \frac{\sigma_\perp B_0^2}{\rho}\Delta h + \frac{\sigma_\perp B_0^2}{\rho}h\Delta h - \frac{\sigma_\perp B_{0y}^2}{\rho}\frac{\partial^2 h}{\partial x^2}$$
$$-\frac{\sigma_\perp B_{0y}^2}{\rho}h\frac{\partial^2 h}{\partial x^2} + \frac{\partial}{\partial y}(\frac{\sigma_\perp B_0^2}{\rho})\frac{\partial h}{\partial y} + \frac{\sigma_\perp B_{0z}^2}{\rho f_0}\frac{\partial}{\partial y}(\frac{\sigma_\perp B_0^2}{\rho})\frac{\partial h}{\partial x}$$
$$+\frac{\partial}{\partial y}(\frac{\sigma_\perp B_0^2}{\rho})h\frac{\partial h}{\partial y} + \frac{\sigma_\perp B_{0z}^2}{\rho f_0}\frac{\partial}{\partial y}(\frac{\sigma_\perp B_0^2}{\rho})h\frac{\partial h}{\partial x} = 0 \quad . \tag{4.38}$$

Further, we will follow so-called "traditional approximation" for the Rossby waves considering not too small latitudes, allowing to consider $\Omega_y = B_{0y} = 0$, and $B_{0z}^2 \to B_0^2$. In addition, we take into account that at heights of F-layer the Pedersen parameter $\sigma_\perp B_{0z}^2/\rho$ does not depend on the $y$-coordinate. Then Eq. (4.38) becomes

$$\frac{\partial \Delta h}{\partial t} - \frac{1}{r_R^2}\frac{\partial h}{\partial t} + \beta\frac{\partial h}{\partial x} - \beta\frac{\sigma_\perp B_0^2}{\rho f}\frac{\partial h}{\partial y} + \frac{\sigma_\perp B_0^2}{\rho}\Delta h$$
$$+\beta h\frac{\partial h}{\partial x} - \beta\frac{\sigma_\perp B_0^2}{\rho f}h\frac{\partial h}{\partial y} + r_R^2 fJ(h,\Delta h) + r_R^2\beta\frac{\partial h}{\partial x}\Delta h - r_R^2\frac{\sigma_\perp B_0^2}{\rho f}\beta\frac{\partial h}{\partial y}\Delta h \quad (4.39)$$
$$-r_R^2\frac{\sigma_\perp B_0^2}{\rho}\frac{\partial h}{\partial x}\frac{\partial \Delta h}{\partial x} - r_R^2\frac{\sigma_\perp B_0^2}{\rho}\frac{\partial h}{\partial y}\frac{\partial \Delta h}{\partial y} + \frac{\sigma_\perp B_0^2}{\rho}(\nabla h)^2 + \frac{\sigma_\perp B_0^2}{\rho}h\Delta h = 0.$$

Here we ignored unnecessary for Rossby waves $h\partial\Delta h/\partial t$ and $\Delta h\partial h/\partial t$ terms [10, 18], and all quantities are taken at the reference point $y = 0$.

It is seen from Eq. (4.39) that linear waves are weakly damped if the dispersion $\beta\partial h/\partial x \geq \sigma_\perp B_0^2\Delta h/\rho$, which gives

$$\frac{\sigma_\perp B_0^2}{\rho f} \leq \frac{L}{R} \sim 10^{-1}. \quad (4.40)$$

Then in Eq. (4.39) we ignore the linear term $\beta\sigma_\perp B_0^2\partial h/\partial y/\rho f$ compared to $\beta\partial h/\partial x$ as small as $L/R$. Further, the nonlinear term $r_R^2\sigma_\perp B_0^2\beta\Delta h\partial h/\partial y/\rho f$ is small as $L/R$ compared to $r_R^2\beta\Delta h\partial h/\partial x$. But, $r_R^2\beta\Delta h\partial h/\partial x$ is small compared to $r_R^2 fJ(h,\Delta h)$ also as $L/R$. Nonlinear terms

$$r_R^2\frac{\sigma_\perp B_0^2}{\rho}\frac{\partial h}{\partial x}\frac{\partial \Delta h}{\partial x} \sim r_R^2\frac{\sigma_\perp B_0^2}{\rho}\frac{\partial h}{\partial y}\frac{\partial \Delta h}{\partial y} \ll r_R^2 fJ(h,\Delta h) \quad (4.41)$$

are small as $L/R$. Also $\beta h\sigma_\perp B_0^2\partial h/\partial y/\rho f \ll \beta h\partial h/\partial x$ is small as $L/R \ll 1$. Thus, Eq. (4.39) can be simplified as follows

$$\frac{\partial \Delta h}{\partial t} - \frac{1}{r_R^2}\frac{\partial h}{\partial t} + \beta\frac{\partial h}{\partial x} + \frac{\sigma_\perp B_0^2}{\rho}\Delta h + r_R^2 fJ(h,\Delta h) + \frac{\sigma_\perp B_0^2}{\rho}(\nabla h)^2$$
$$+\beta h\frac{\partial h}{\partial x} + \frac{\sigma_\perp B_0^2}{\rho}h\Delta h = 0 \quad (4.42)$$

Thus along with known KdV nonlinearity $\beta h\partial h/\partial x$ term two nonlinear scalar terms appeared in the nonlinear Eq. (4.42). They contribute with vector nonlinearity $r_R^2 fJ(h,\Delta h)$ if

$$\frac{r_R^2}{L^2} \sim \frac{L}{R} \sim \frac{\sigma_\perp B_0^2}{\rho f} < 1, \quad (4.43)$$

i.e. in case of large-scale structures, namely when $L \sim (r_R^2 R)^{1/3}$. This means that radius of structure $L$ should be more than so-called intermediate geostrophic radius $r_{ig} = r_R^{2/3} R^{1/3}$.

Eq. (4.42) conserves mass density (see Eq. (4.8)), and the $y$-coordinate for the center of mass (see Eq. (4.11)). As to the energy, instead of Eq. (4.9) we get

$$\frac{\partial}{\partial t}\int \varepsilon dxdy = -\frac{\sigma_\perp B_0^2}{\rho} r_R^2 \int dxdy(\nabla h)^2 - \frac{\sigma_\perp B_0^2}{\rho} r_R^2 \int dxdy h(\nabla h)^2 . \tag{4.44}$$

Thus, the total energy in the F-layer is not conserved and damped due to the Pedersen conductivity. As to the potential enstrophy instead of Eq. (4.12) we get

$$\frac{\partial}{\partial t}\int dxdy[(\nabla h)^2 + r_R^2 (\Delta h)^2] = -2r_R^2 \frac{\sigma_\perp B_0^2}{\rho}\int dxdy(\Delta h)^2 - r_R^2 \beta \int dxdy \Delta h \frac{\partial h^2}{\partial x}$$
$$-r_R^2 \frac{\sigma_\perp B_0^2}{\rho}\int dxdy h^2 \Delta^2 h \tag{4.45}$$

Here

$$(\Delta h)^2 = (\frac{\partial^2 h}{\partial x^2})^2 + 2\frac{\partial^2 h}{\partial x^2}\frac{\partial^2 h}{\partial y^2} + (\frac{\partial^2 h}{\partial y^2})^2$$
$$\Delta^2 h = \frac{\partial^4 h}{\partial x^4} + 2\frac{\partial^4 h}{\partial x^2 \partial y^2} + \frac{\partial^4 h}{\partial y^4} \tag{4.46}$$

Thus, the potential enstrophy $Q = \langle (\nabla h)^2 + r_R^2(\Delta h)^2 \rangle$ is not conserved due to Pedersen conductivity and scalar nonlinearity. Note that the last two terms on the right-hand side is small as $O(h^3)$.

## V. Discussion and Conclusion

In this study, in the frame of shallow water model we revealed the influence of free surface action on nonlinear propagation of planetary ULF magnetized Rossby waves in the Earth's weakly ionized ionospheric D-, E-, and F-layers. Ionospheric charge particles electrodynamic effects are taken into account by inclusion of Hall and Pedersen conductivities (see Eqs. (2.5) and (2.6)). Corresponding nonlinear equation valid for D-, E-, and F-layers is obtained (see Eq. (4.4). The main result of the free surface action is the appearance of the scalar (Korteweg-de Vries (KdV) type) nonlinearity $\sim \beta h \frac{\partial h}{\partial x}$ in the dynamic equations (see Eqs. (4.7), (4.19), (4.42)). In addition, Eq. (4.42) reveals existence of two new scalar nonlinearities $\sim (\nabla h)^2, h\Delta h$ due to Pedersen conductivity in the ionospheric F-layer. These three scalar nonlinearities contribute in the dynamical formation of large-scale vortical structures when their size $L \sim (r_R^2 R)^{1/3}$. Thus, the dynamics of magnetized Rossby waves in general is stimulated by both vector (Jacobian) and scalar nonlinearities. Note, scalar nonlinearity arises from the free-surface $h$ variation, $\beta$-effect and Pedersen parameter $\sigma_\perp B_0^2 / \rho$. The other vector nonlinearity, $\sim J(h,\Delta h)$, arises from the convective derivative of the $z$-component of the vorticity $\mathbf{v}\cdot\nabla\zeta_z$, and always occurs for a rotating fluid, even with a rigid surface. Nonlinear vorticity equations governing the evolution of planetary ULF magnetized Rossby waves in the Earth's ionospheric D-, E-, and F-layers obtained (see Eqs. ((4.7), (4.19), (4.42)). Expressions for the potential vorticity defined for D-, E-, and F-layers (see Eqs. (3.5) and (3.7)). It is shown that while potential vorticity is conserved in the ionospheric D-, and E-layer (see Eqs. (3.4) and (3.6)), it is broken by Pedersen conductivity in the ionospheric F-layer (3.8). It means that magnetized Rossby waves are damped in the F-layer. Obtained dynamic equations (4.7), (4.19), and (4.42) satisfy mass, energy, $y$ - coordinate of center of mass, and potential enstrophy conservation laws (see Eqs. (4.8)-(4.13), (4.20)-(4.21), (4.44)-(4.46)).

Finding of main characteristic nonlinear solutions is the subject of our next investigation.

Obtained in the given investigation results extend and complement known theoretical investigations in the geophysical fluid mechanics and are especially relevant for nonlinear vortical propagation of magnetized Rossby waves in the weakly ionized ionospheric plasma.

**Data Availability Statement**

Data sharing is not applicable to this article as no new data were created or analyzed in this styudy.